# Klein – Gordon equation for market wealth operations

Magdalena Pelc

Institute of Physics, Maria Curie Sklodowaka University, Lublin, Poland

Abstract

In this paper the modified Klein – Gordon equation for market processes is proposed and solved. It is argued that the oscillations in market propagate with the light velocity. The initial pulse in the market is damped and for very large time diffused according to the Fourier law.

Introduction

In recent years, easy availability of market data has helped in the analysis of wealth or income distribution in various societies [1].

There has been several attempts to model a simple economy with minimum trading ingredients, which involve a wealth exchange processes [2] that produce a distribution of wealth similar to that observed in the real market. In this paper we are particularly interested in microscopic models of markets where the trading activity is considered as a scattering process [3].

In the paper following the idea of the modeling of market as an ideal gas model we develop the more complex description which leads to the Klein – Gordon equation for the economics processes.

1. The model

All gas – like models of trading market are based on the assumptions: (a) money conservation (globally in the market; as well as locally in any trading) and (b) stochasticity. In the view the trading as scattering processes one can see the equivalence. If we concentrate only on the cash exchanged (even using bank cards) every trading is a money conserving one – like elastic scattering processes in physics. In this paper we allow the change of number of peoples (partners) which take a part in market processes.

Let $P(x,t)$ denotes the number density of people processing wealth and active in the market.

In the subsequent we assume the correlated random walk model for change of the $P(x,t)$. In that case $P(x,t)$ fulfils the Heaviside equation [4]

$$\tau \frac{\partial^2 P(x,t)}{\partial t^2} + \frac{\partial P}{\partial t} = D \frac{\partial^2 P}{\partial x^2}, \qquad (1)$$

where $\tau$ denotes the relaxation time for market processes, $D$ is the diffusion coefficient. The Eq.(1) can be solved with the substitution

$$P(x,t) = e^{-1/2\tau} u(x,t). \qquad (2)$$

From formulae (1) and (2) we obtain for $u(x,t)$

$$\frac{1}{v^2} \frac{\partial^2 u(x,t)}{\partial t^2} - \frac{\partial^2 u(x,t)}{\partial x^2} + q^2 u(x,t) = 0$$

$$q^2 = -\left(\frac{v}{D}\right)^2 \qquad (3)$$

Equation (3) is the modified Klein – Gordon equation for market processes.
From statistical physics it is well known that

$$D = \tau v^2 \qquad (4)$$

and formula (4) can be written as

$$q^2 = -\left(\frac{1}{\tau v}\right)^2. \qquad (5)$$

In formula (5) $L_D = \tau v$ denotes the mean free path for the agent interaction.

Equation (1) can be regarded as the approximation to the Fokker – Planck equation developed for market processes by R. Friedrich et al. [5]. For $t \to \infty$ equation (1) is the diffusion Fourier equation for $t/\tau \gg 1$ the first term in Eq.(1) can be omitted.

For Cauchy problem, i.e. for

$$u(x, t = 0) = f(x),$$
$$\frac{u(x, t = 0)}{dt} = g(x). \tag{6}$$

The general solution of the Klein – Gordon equation has the form

$$u_m(x,t) = \frac{f(x-vt) + f(x+vt)}{2}$$
$$+ \frac{1}{2v} \int_{x-vt}^{x+vt} g(\varsigma) I_0 \left[ \sqrt{-q^2 \left(v^2 t^2 - (x-\varsigma)^2\right)} \right] d\varsigma \tag{7}$$
$$+ \frac{v\sqrt{-q^2}\, t}{2} \int_{x-vt}^{x+vt} f(\varsigma) \frac{I_1 \left[ \sqrt{-q^2 \left(v^2 t^2 - (x-\varsigma)^2\right)} \right]}{\sqrt{v^2 t^2 - (x-\varsigma)^2}} d\varsigma$$

In Eq.(7) $I_0$ and $I_1$ are the modified Bessel functions.

2. Elementary processes in market operation

Let us consider equation (3) in more details. First of all for $q \to 0$ Eq.(3) is the well known wave equation:

$$\frac{1}{v^2} \frac{\partial^2 u}{\partial t^2} - \frac{\partial^2 u}{\partial x^2} = 0 \tag{8}$$

with general solution

$$u(x,t) = \frac{f(x-vt) + f(x+vt)}{2}. \tag{9}$$

Eq.(9) describes the wave propagation with velocity $v$ and $P(x,t)$ can be written as

$$P(x,t) = e^{-\frac{1}{2\tau}} \left[ \frac{f(x-vt) + f(x+vt)}{2} \right] \tag{10}$$

i.e. Eq.(1) is the damped wave equation.

The initial oscillation of the market in place X propagates through the full population of the agents with velocity $v$. As in contemporary economics information is transferred by electromagnetic field, $v = c = 3 \cdot 10^8$ m/s.

For $\tau \to 0$ equation (1) is the Fourier diffusion equation

$$\frac{\partial P}{\partial t} = D \frac{\partial^2 P}{\partial x^2} \tag{13}$$

and initial oscillation of the market is very quickly damped. Instead of wave we obtain the diffusion.

Conclusions

In this paper we analyze the market processes as the correlated random walk processes. We argue that the market operations fulfill the modified Klein – Gordon equation (MKG). The solution of the MKG consists of the two components: wave and wave drag. For very long time market processes can be described by the Fourier type diffusion equation.